\documentclass[prb,preprint,showpacs,preprintnumbers,amsmath,amssymb]{revtex4}
\usepackage{graphicx}
\usepackage{amssymb}
\usepackage{mathrsfs}
\usepackage{dcolumn}
\usepackage{bm}
\newcommand{\lmo}{LaMnO$_3$ }
\newcommand{\eg}{$e_g$ }
\newcommand{\tg}{$t_{2g}$ }
\newcommand{\J}{$J_1$ }
\newcommand{\jj}{$J_2$ }
\newcommand{\tjt}{$T_{JT}$}
\newcommand{\dx}{$d_{3x^3-r^2}$ }
\newcommand{\dy}{$d_{3y^3-r^2}$ }
\newcommand{\dz}{$d_{3z^3-r^2}$ }

\begin{document}
\title{A Potts model for the distortion transition in LaMnO$_3$ }

\author{Mahrous R. Ahmed\footnote{Permanent address:  Department of Physics, Faculty of Science, South Valley University, 82524 Sohag, Egypt.}}
\email[Corresponding author: ]{php02mra@shef.ac.uk,  mahrous_r_ahmed@yahoo.com}
\author{G. A. Gehring}

\affiliation{Department of Physics and Astronomy, University of Sheffield, Hicks Building, Hounsfield Road, Sheffield, S3 7RH, UK.}

\date{\today}

\begin{abstract}
The Jahn-Teller distortive transition of \lmo is described by a modified 3-state Potts model. The interactions between the three possible orbits depends both on the orbits and their relative orientation on the lattice. Values of the two exchange parameters which are chosen to give the correct low temperature phase and the correct value for the transition temperature are shown to be consistent with microscopy theory. The model predicts a first order transitions and also a value for the entropy above the transition in good agreement with experiment. The theory with the same parameters also predicts the temperature dependence of the order parameter of orbital ordering agreeing well with published experimental results. Finally, the type of the transition is shown to be close to one of the most disordered phases of the generalised Potts model. The short range order found experimentally above the transition is investigated by this model.
\end{abstract}


\maketitle
\parindent1cm

\section{\label{intro}Introduction}
Orbital ordering and Jahn-Teller (JT) distortions are observed in some perovskite-type 3d transition-metal compounds such as the vanadates (e.g. V$_2$O$_3$,\cite{Castellani78} LiVO$_3$,\cite{Pen97} and LaVO$_3$.\cite{Khaliullin01}) cuprates\cite{Kugel82} (e.g. KCuF$_3$) and derivatives of the colossal magnetoresistive in the compound \lmo,\cite{Brink02,Murakami98} which is the subject of this paper. 

A single electron, or hole, in cubic symmetry will be in a two-fold or three-fold degenerate state if it occupies a \eg or \tg orbit. This degeneracy will be lifted at a transition where the local symmetry is lowered and long range orbital order occurs in the low temperature-phase\cite{Wollan55,Rdriguez98}. The local orbital order will be accompanied by local structure distortions because of a linear coupling between the occupations of the different orbital states and the lattice\cite{Khomskii2003}. This is the JT effect. In a perovskite structure this is a distortion and rotation of the oxygen octahedra where in the lattice structure ABO$_3$ (where A= rare-earth cation and B=Mn) the B atom is octahedrally coordinated to the oxygen. The local distortion around one B ion affects other ions in the crystal because the oxygen ions are shared between two octahedra. If a large fraction of the energy of the orbital ordering comes from the energy lowering due to the interaction with the lattice the interaction is known as a cooperative JT effect\cite{Millis96}. There is a linear coupling between an \eg orbital and the E distortion modes of the octahedron. The coupling may be determined from a classical harmonic approximation to the lattice dynamics\cite{Millis96}. 

The subject has attracted much interest recently because of the interplay between the magnetism and the lattice. The coulomb repulsion alone which gives rise to magnetic superexchange\cite{Khaliullin01} can also drive orbital order and hence there has been much efforts to determine the relative importance of the lattice effects, cooperative JT, and the electronic energies in driving the transition. 

\begin{figure}[h!]
\begin{center}
\includegraphics[scale=0.5]{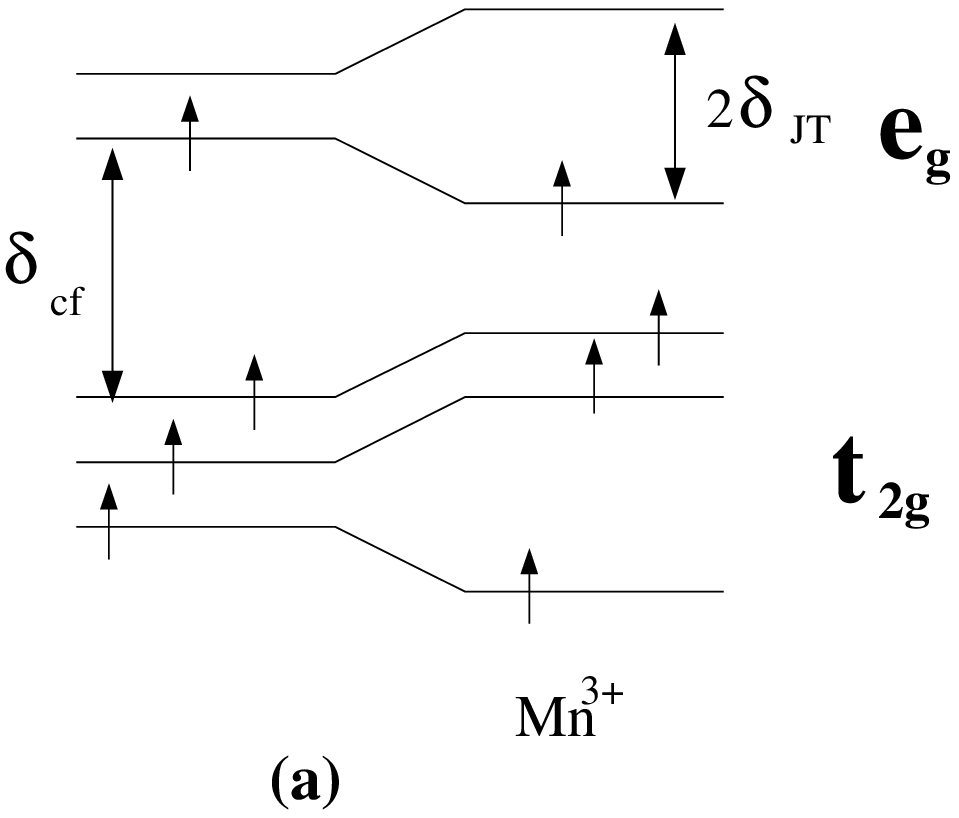}\hspace{1.0cm} \includegraphics[scale=0.4]{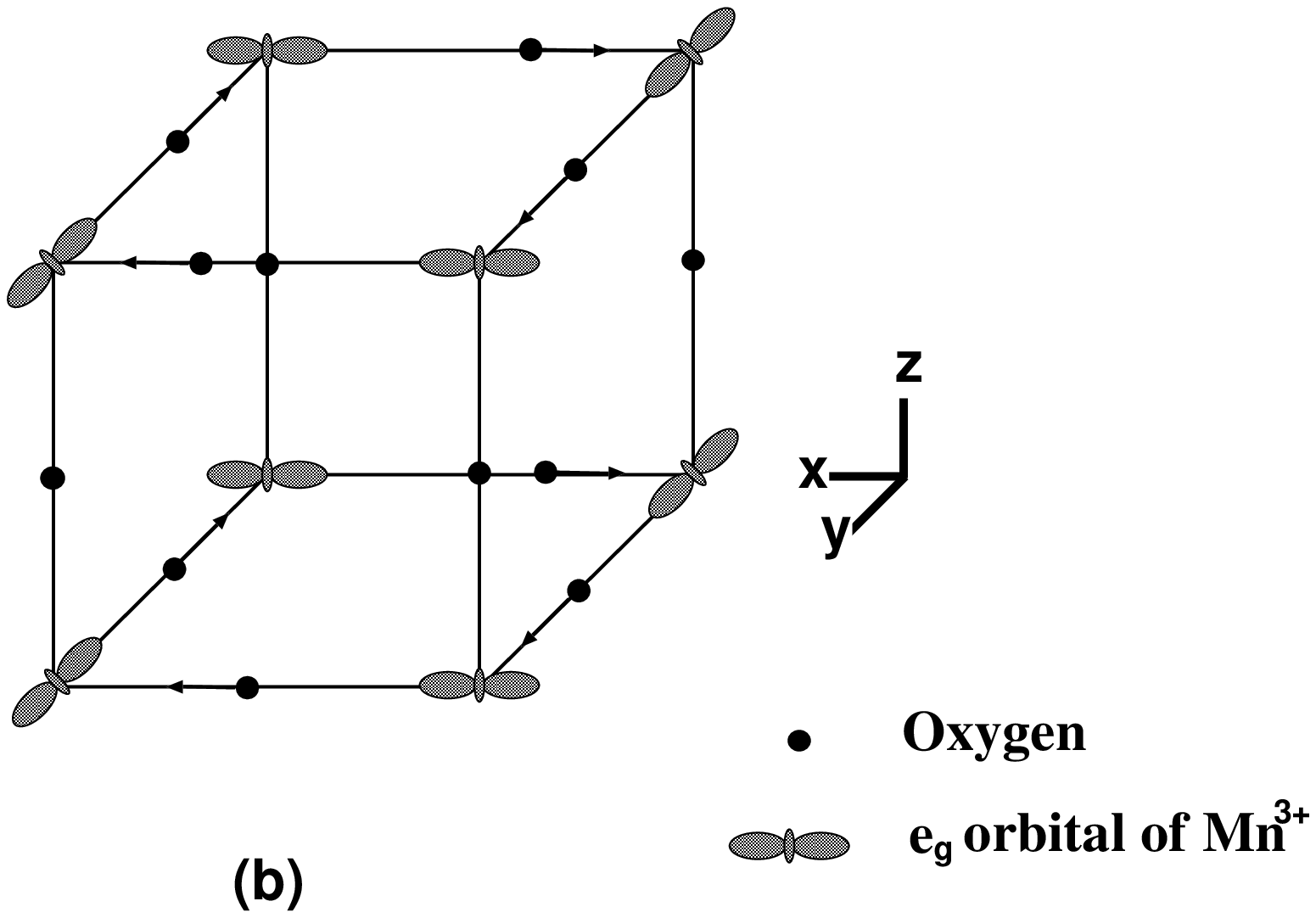}
\end{center}
\caption{\label{oo}(a) Splitting of 3d-levels in a cubic crystal field with an additional tetragonal distortion. Electron occupation of the four 3d-electrons in Mn$^{3+}$ is shown by arrows, (b) the orbital ordering of the \eg states in an octahedral crystal of \lmo at $T\le T_{JT}$. One of the two bonds of Mn-O-Mn bond in the $x-y$ plane contracts and the other expands. The La$^{3+}$ ion in the centre of the cell is omitted for clarity}
\end{figure}
We consider \lmo, where Mn$^{3+}$ has a $d^4$ configuration, i.e., four electrons in $d$ orbitals. The wave functions, $d_{x^2-y^2}$ and \dz, are called $e_g$ orbitals, whereas, $d_{xy}$, $d_{yz}$, and $d_{zx}$ are called $t_{2g}$ orbitals. The level splitting is shown in Fig. \ref{oo}a in a cubic crystal and with a tetragonal distortion. The orbital ordering is shown  in figure \ref{oo}b. When electrons occupy these wave functions, the ground state is determined by the semiempirical Hund's rule in \lmo because the exchange energy is greater than the crystal field energy, $\delta_{cf}$. Hund's rule coupling causes all of the spins of the electrons to be aligned parallel, that is $S=2$, where three electrons are in the $t_{2g}$ orbitals and one electron occupies one of the $e_g$ orbitals. The MnO$_{6}$ octahedra in the low temperature phase have two short, two medium, and two long bonds\cite{Proffen99} while above \tjt \@ the octahedra have six almost equal bond lengths.   

This paper develops a model for the temperature dependence of the orbital ordering. There have been several studies of the effect of temperature in the orbitally ordered phase. The orbital ground state is well known as a C-type, antiferromagnetic in x-y plane and ferromagnetic in z-direction, with long-range ordered, JT distorted, MnO$_6$ octahedra\cite{Wollan55,Rdriguez98} as sketched in Fig. \ref{oo}b. The elongated occupied \eg orbitals alternate between pointing along x and y directions, the so-called orthorhombic structural phase\cite{Wollan55,Rdriguez98}. Above \tjt, the structural phase is pseudocubic with almost regular MnO$_6$ octahedra. Also above \tjt \@ the long-range orbital ordering disappears. The structure was studied using the thermal analysis high-resolution neutron-powder diffraction patterns\cite{Rdriguez98} and Raman phonons as function of temperature\cite{Carron01}. Evidence has also been presented\cite{Murakami98} for orbital ordering in \lmo which is based on the splitting of the Mn 4p levels by the Mn 3d orbital ordering. 

\lmo undergoes a structural phase transition\cite{Rdriguez98,Okamoto02} at the JT temperature, \tjt \@ $= 750 K$. Murakami {\it et al.}\cite{Murakami98} showed that the order parameter of orbital ordering as a function of temperature decreases rapidly for $T\ge0.4$\tjt \@ and disappears at $T_{JT}=780 K$ in a first order transition. These results have been re-interpreted and fitted\cite{Deisenhofer03} by a critical exponent, $\beta=0.16(1)$, at \tjt \@ when the temperature dependence of the electron-spin resonant in La$_{0.95}$Sr$_{0.05}$MnO$_3$ has been investigated.
 
S\'anchez {\it et al}.\cite{Sanchez03} showed by means of X-ray absorption near edge structure and extended X-ray absorption fine structure at the Mn $K$-edge that the structural transition originates by the ordering of tetragonally distorted octahedra and they analysed the entropy content of the transition within the framework of the three-state Potts model\cite{Potts52,Wu82,Wang90} with nearest-neighbour antiferrodistortive coupling. It is confirmed\cite{Xiangyun05}, by using neutron diffraction data and a combined Rietveld and high real space resolution atomic pair distribution function analysis, that the nature of the JT transition around \tjt \@ is to be orbital order-disorder and the intermediate structure suggests the presence of local ordered clusters implying strong nearest-neighbour JT antidistortive coupling.

Millis\cite{Millis96} derived a classical model, which was based on previous work by Kanamori\cite{Kanamori61},  for the lattice distortions in manganites. The model may be approximated either by an antiferromagnetic $xy$ model with a modest threefold anisotropy or by a 3-state Potts model with an antiferromagnetic first-neighbour interaction and a weak second-neighbour interaction. This differs from our model which deals only with the nearest neighbour interaction. 

We study the orbital ordering appropriate for \lmo obtained from our model and the JT distortion which occurs above JT temperature in that phase. The model proposed here for the disordered phase is an array of localised distortions with no long range order. This differs from the model of Zhou-Goodenough\cite{Zhou99} who postulated that dynamic JT effects occur in the high temperature phase. Zhou and Goodenough suggested that the orbital order-disorder transition at \tjt \@ is to a dynamic JT stabilisation at the $e_g$ electrons persisting into the high-temperature O-orthorhombic phase, and they proposed a vibronic mechanism to explain the electrical resistivity and thermoelectric power above and below \tjt \@ respectively. 

It is well known that the phenomenological models such as pseudo spin or Potts models are very useful to study order-disorder transitions. In the \lmo system, the MnO$_6$ octahedra are not independent as they share oxygen atoms with their nearest neighbours. The standard 3-state Potts model alone does not have the correct orbital ordering for \lmo as a possible ground state. In this paper, we describe a phenomenological model that gives the correct ground state for the orbital ordering and a good value for high-$T$ entropy for the \lmo phase. The model is called the anisotropic Potts model and has been presented earlier\cite{Ahmed05}. 

In section \ref{Methodology}, the anisotropic Potts model and its phase diagram are presented and its application to \lmo is in section \ref{LMO}. The model is used to give understanding of the orbital ordering just above \tjt \@ in section \ref{J2=0}. Finally, we conclude the results in the last section.

\section{\label{Methodology}Methodology}
\subsection{\label{model}The model}
We model the transition at \tjt \@ considering only the largest distortions that leave the crystal tetragonal in the low temperature phase. If the tetragonal axis is chosen along $\hat{z}$ direction the orbital ordering is characterised by the vector ${\bf Q}=(110)$. There is a staggered order in the $x-y$ plane and the orbits are stacked ferromagnetically up the $\hat{z}$ axis as shown in Fig. \ref{oo}(b). However, in Fig. \ref{oo}(b) we showed the ordering for a pure JT system where the orbits shown were $|x\rangle=|3x^2-r^2\rangle$ and $|y\rangle=|3y^2-r^2\rangle$. These orbits are found from the \eg doublet for the case $\theta = \pm \frac {2\pi}{3}$.
\begin{equation}
|\theta\rangle=cos\frac{\theta}{2}|3z^2-r^2\rangle+sin\frac{\theta}{2}|x^2-y^2\rangle.
\end{equation}
Experimentally\cite{Deisenhofer03} it is found that if the tetragonal axis is along  $\hat{z}$ direction, then, the staggered orbits are given by $|\pm\theta_{exp}\rangle$ where $\theta_{exp}$ is closer to $\pi/2$ than $2\pi/3$. We note that,
\begin{equation}
|\pm \frac{\pi}{2}\rangle=cos\frac{\pi}{4}|3z^2-r^2\rangle+sin\frac{\pi}{4}|x^2-y^2\rangle.
\end{equation}
This gives, when $|-\frac{\pi}{2}\rangle$ is used, $\frac{1}{\sqrt{6}}|2.7x^2-0.7y^2-2z^2\rangle$ which has its main lobe along $\hat{x}$ direction. Similarly, using $|+\frac{\pi}{2}\rangle$ gives a state, $\frac{1}{\sqrt{6}}|2.7y^2-0.7x^2-2z^2\rangle$, which has its main lobe along $\hat{y}$ direction. We use the JT states in our phenomenological model because these allow us to use the same states below \tjt \@ for any of the three possible ${\bf Q}$ vectors, (110), (101) and (011), and also above \tjt. The model presents the correct dimension of the order parameter. 

There are six equivalent orderings that can occur. These correspond to the three choices for the tetragonal axis and, then, the phase ($\pm1$) of the order parameter. In the high temperature phase of the Potts model the orbits are in one of states $\theta=0$ or $\theta=\pm\frac{2\pi}{3}$. This differs from Zhou and Goodenough\cite{Zhou99} who assumed that above \tjt \@ the orbit is rotating, thus, all values of $\theta$ are accessed dynamically. Because the orbital ordering in the transition metal oxides has three anisotropic states it is natural to set up a three states Potts model. The standard $q$-state Potts model\cite {Potts52} consists of a lattice of spins , which can take $q$ different values from 1 to $q$, and whose Hamiltonian is
\begin{equation}
{\mathscr{H}}=-\frac{J}{2}\sum_{\langle i,j\rangle}^{N}\delta_{S_{i},S_{j}},
\label{iso}
\end{equation}
where $S_{i}=1,2,$..... is one of the $q$ states on site $i$, ${\delta}_{S_{i},S_{j}}$ is the Kronecker function which is equal to 1 when the states on sites $i$ and $j$ are identical, $S_i=S_j$, and is zero otherwise, $\langle i,j\rangle$ means that the sum is over the nearest neighbour pairs, $J$ is the exchange integral and $N$ is the total number of sites in the lattice. For $q=2$, this is equivalent to the Ising model. The Potts model is, thus, a simple extension of the Ising model, however, it has a much richer phase structure, which makes it an important testing ground for new theories and algorithms in the study of critical phenomena\cite{Wu82}.

\begin{figure}[h!]
\includegraphics[scale=0.45]{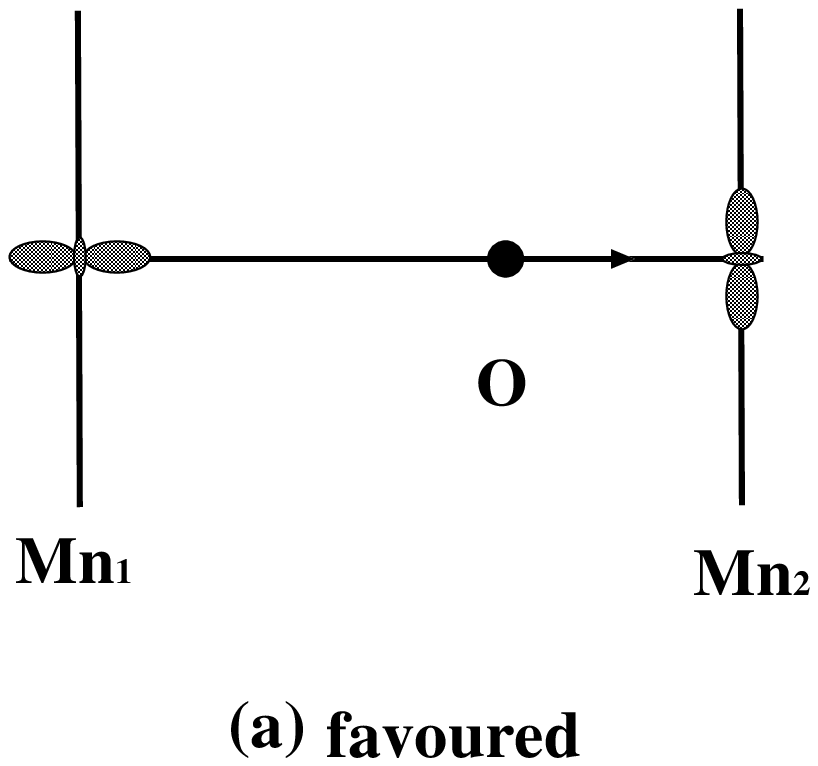}\hspace{1cm}
\includegraphics[scale=0.45]{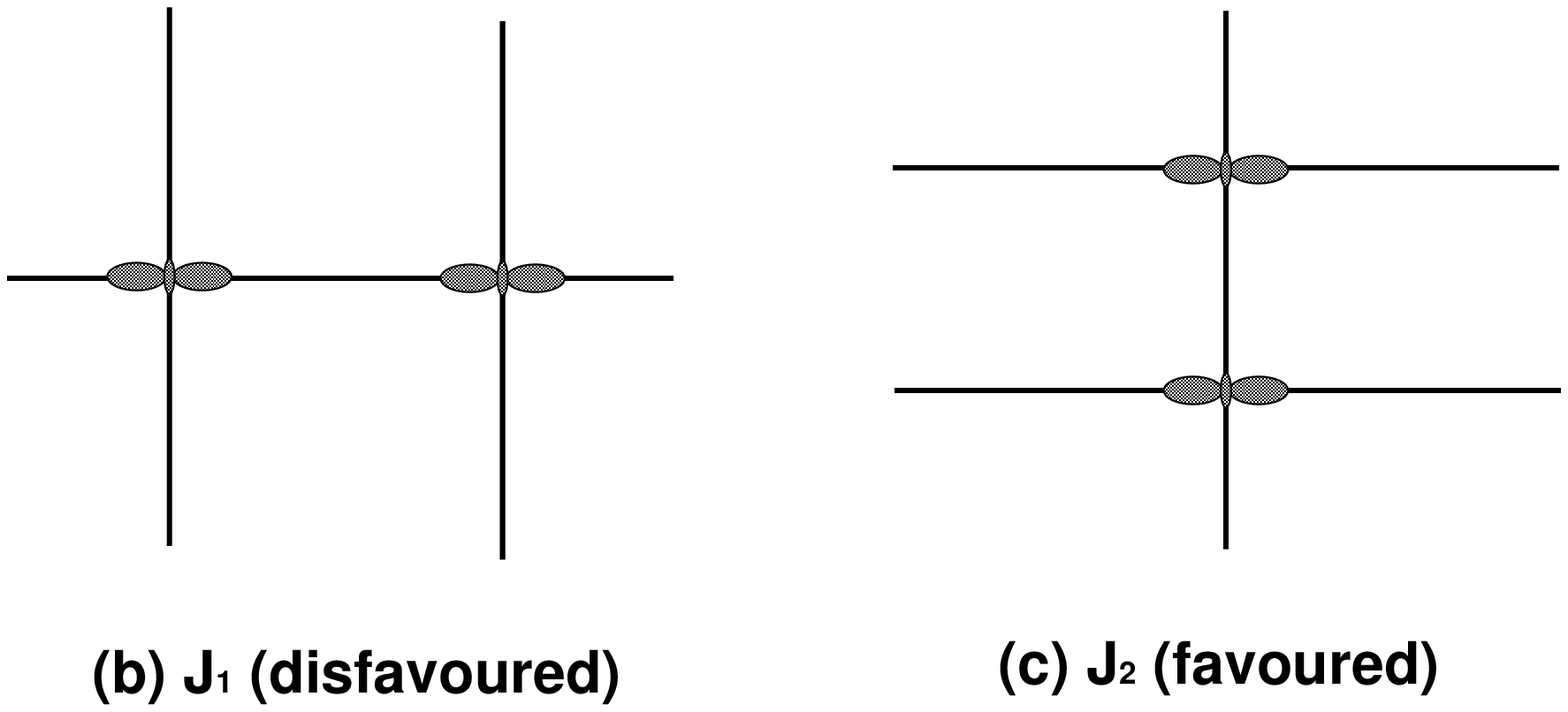}
\caption{\label{J1}The exchange interactions used in the anisotropic Potts model to give orbital ordering in \lmo, (a)  if a Mn$_{1}$-O bond expands, the oxygen shared O-Mn$_{2}$ bond should contract, (b) Two identical orbits in the same lattice vector, $J_1>0$, disfavoured, (c) Two identical orbits in different lattice vector, $J_2=0$, favoured.}
\end{figure}
We have presented\cite{Ahmed05} a modified Potts model on a cubic lattice where the interaction between two sites depends on both the orbitals on these sites and also the directions of the nearest neighbour bond between them. We have found that this general model describes the order seen in \lmo for a range of parameters. In \lmo, as we explained above, the MnO$_{6}$ octahedra are not independent as they share oxygen atoms with their nearest neighbours (see Fig. \ref{J1}(a)). This introduces a constraint in the degrees of freedom for each octahedron \cite{Sanchez03}. If a Mn$_{1}$-O bond expands, the oxygen sharing O-Mn$_{2}$ bond should contract and vice versa. According to that, Mn orbital ordering is not isotropic. Namely, if Mn-O-Mn was located along $\hat{x}$ direction in $x-y$ plane, the $3d$ orbit of Mn in the expanded Mn$_{1}$-O bond will be \dx, $x$-state, but in the contracted O-Mn$_{2}$ bond it could be either \dy, $y$-state, or \dz, $z$-state, and the same order may exist in the next sublattice, as shown in Fig. \ref{oo}(b). The energy of a  pair of identical orbits is \J only if the lattice vector joining them is along the lobe of the orbit as shown in Fig. \ref{J1}(b). The interaction energy of two identical orbits separated by a lattice vector that is not along the lobe is \jj as shown in Fig. \ref{J1}(c). In 3-state 3-dimensions anisotropic  Potts model each site is occupied by one of the orbits $x$, $y$ or $z$ and the energy is described by the following Hamiltonian,
\begin{equation}
{\mathscr H_{aniso}}=-\frac{1}{2}\sum_{\langle i,j\rangle}^{N}J_{S_i} (\underline {\rho}_{ij}) \delta_ {S_{i}, S_{j}},
\label{aniso}
\end{equation}
where $\underline{\rho}_{ij}=\underline{R}_i-\underline{R}_j$ is the direction of the nearest neighbour bond between two states.

There are two exchange interactions for the anisotropic model. The `head to head' interaction $J_1$ is defined by $J_{x}(\underline{\rho}) = J_1$ for $\underline{\rho}=\pm \hat{x}a$, $J_{y}(\underline{\rho}) = J_1$ for $\underline{\rho}=\pm \hat{y}a$ and $J_{z}(\underline{\rho}) = J_1$ for $\underline{\rho}=\pm \hat{z}a$. The `side to side' interaction $J_2$ is defined by $J_{x}(\underline{\rho})=J_2$ for $\underline{\rho}=\pm \hat{y}a$ or $\underline{\rho}=\pm \hat{z}a$, $J_{y}(\underline{\rho})=J_2$ for $\underline{\rho}=\pm \hat{x}a$ or $\underline{\rho}=\pm \hat{z}a$ and $J_{z}(\underline{\rho})=J_2$ for $\underline{\rho}=\pm \hat{x}a$ or $\underline{\rho}=\pm \hat{y}a$. Thus each site has a coupling $J_2$ to four neighbours and a coupling $J_1$ to two neighbours. This is shown in Fig. \ref{J1}(b) and \ref{J1}c. It is worth mentioning that these types of interaction do not affect the overall cubic symmetry of the lattice. 
\subsection{Monte Carlo Simulations}
This model has been studied using Monte Carlo (MC) simulations on 3-d finite lattices (with linear sizes $L$ = 8, 10 and 12) for a range of values of \J and \jj with periodic boundary conditions. All our simulations have made use of the Metropolis algorithm and with averaging performed over from $10^5$ to $10^7$ Monte Carlo steps per site. Results were obtained by either cooling down from a high-temperature random configuration as discussed by Banavar {\it et al}.\cite{Banavar82} or heating up from the ground state. The results from the two procedures agree. 

\begin{figure}[hi]
\includegraphics{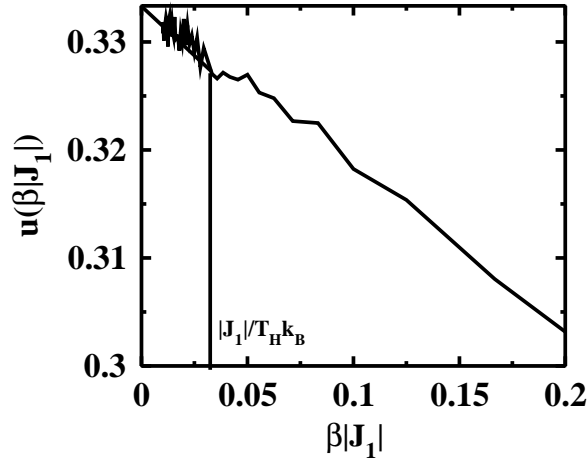}
\caption{\label{TH}Energy, $u$, per site versus $\beta|J_1|$ (where $\beta=1/k_BT$) for the anisotropic 3-state 3-d Potts model with $L=8$ in the low $\beta$ limit. The suitable starting point for our simulation is obtained, $T_H\simeq32|J_1|/k_B$. For simplicity we show the case where $J_2=0$.}
\end{figure}
Combination of the analytic method\cite{Binder81} in the range $T_H<T<\infty$ and computational results in the range $T<T_H$ can be used to obtain the high-$T$ limit to get the starting point, $T_H$, for our simulations. The high-$T$ expansion of the internal energy as a function of temperature, $u(T)$, from the Hamiltonian in Equation \ref{aniso} for the case $|J_1|>>J_2$ is,
\begin{equation}
u(T)= u(\infty)-\frac{1}{2k_BT_H}\left(\frac{z'J_1^2}{4}\right ),
\end{equation}
where $z'$ is the number of the similar nearest neighbours, $z'=2$, and $u(\infty)=\frac{J_1-2J_2}{3}$. The value of $u(T)$ can be written as follow,
\begin{equation}
u(T) =\frac{J_1-2J_2}{3}-\frac{A}{T_H},
\end{equation}
where $A$ is a constant. Fig. \ref{TH} shows a plot of $u(T)$ in units of \J \@ against $\beta|J_1|$. For small $\beta|J_1|$, $\beta|J_1|=\frac{|J_1|}{k_BT_H}\le0.03125$, then, $T_H\simeq32|J_1|/k_B$. 
\begin{figure}[h!]
\includegraphics[scale=0.4]{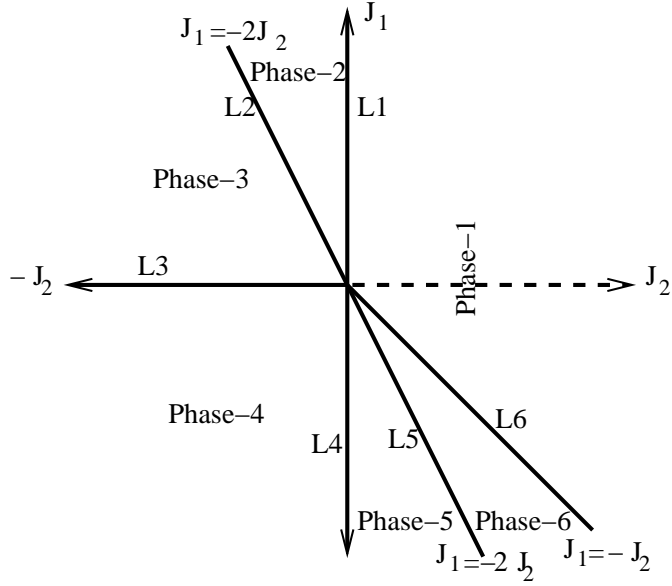}
\caption{\label{phasediag}\J - \jj phase diagram\cite{Ahmed05} of the orbital structures in a simple cubic lattice for general Hamiltonian (Eq. \ref{aniso}). The ground state of \lmo occurs when \J is AF, \jj is FM and $0<\frac{J_2}{|J_1|}<0.5$ which is phase-5 above.}
\end{figure}

We use the simulations to evaluate the thermodynamic quantities in the range $0<T<T_H$, and find the contribution for $T>T_H$ analytically. A full phase diagram with six regional phases in the \J-\jj plane has been obtained numerically and analytically\cite{Ahmed05}, see Fig. \ref{phasediag}. 

\section{\label{LMO}Simulation for $LaMnO_3$}
The phase of interest in Fig. \ref{phasediag} is phase-5 where $J_1<0$ so that the head-to-head configuration is disfavoured and \jj$>0$, so that, side-to-side configuration is favoured and $0<\frac{J_2}{|J_1|}<0.5$. This has the correct low temperature pattern of orbital ordering seen in \lmo, as shown in Fig. \ref{oo}(b). In this phase we have an $x,y$ checkerboard pattern in the $x-y$ plane and this repeats itself so that the $x$ and $y$ states are in ordered chains up the $\hat{z}$ axis. The contribution to the energy comes from these ferromagnetic chains so the ground state energy is given by $u(0)=-J_2$. We find that at high-$T$ the internal energy is $u(\infty)=\frac{1}{3}(J_1-2J_2)$ as expected from a random orbital array in the Potts model. The stabilisation energy of this phase is obtained as, 
\begin{equation}
\Delta u=u(\infty)-u(0)=\frac{1}{3}(J_1-J_2).
\end{equation}

\vspace{0.5cm}
\begin{figure}[hi]
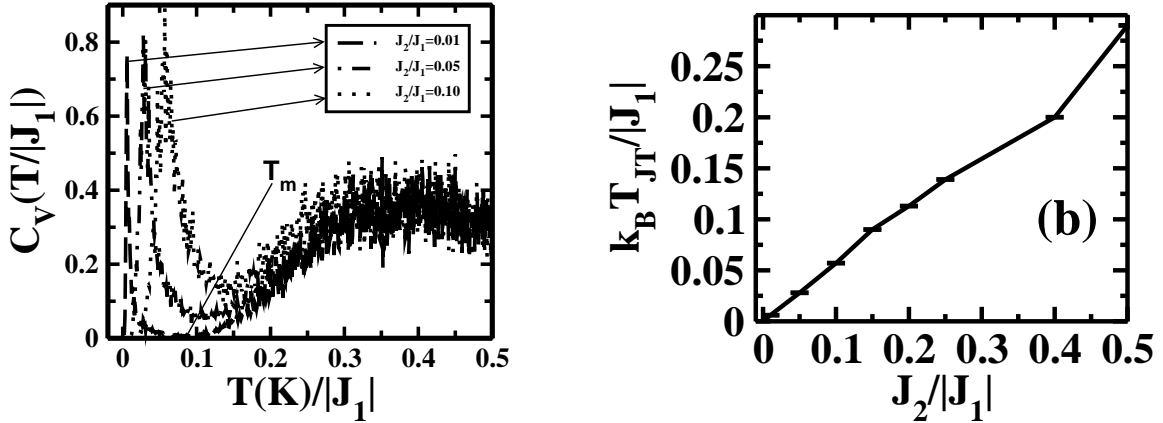

\includegraphics[scale=0.6]{Cv_T_-0.01_-0.05_-0.1.eps} \hspace{1.0cm} \includegraphics[scale=0.75]{Tc_J2.eps}
\caption{\label{Cv}(a) Temperature dependence of specific heat, $C_V$, per site for the 3-state 3-d anisotropic Potts model at $L=12$ where $J_2/|J_1|=0.01, 0.05$, and 0.1 and \J$<0$. $T_m$ is the temperature at the minimum value of $C_V$ between the two peaks. (b) the dependence of the transition temperature on $J_2/|J_1|$.}
\end{figure}
The behaviour of the specific heat, $C_V(T/|J_1|)$, in \lmo phase produced by this model (see Fig. \ref{Cv}(a)) shows that it has a sharp peak corresponding to a first order transition in the range $0<J_2/|J_1|<0.5$ and in the range $0<J_2/|J_1|<0.25$ there is a second, broader, peak. In the range $0.25<\frac{J_2}{|J_1|}<0.5$ the two peaks merge. The first peak obtained along the whole region of \lmo phase is at the transition from the orbital order to orbital disorder configuration occurring above \tjt. The second peak obtained along the region $0<J_2/|J_1|<0.25$ is caused by the short range order occurring because of the large value of the head-to-head interaction, $J_1$. We identify the short range order below the broad peak with the short range order observed experimentally just above \tjt. In section \ref{J2=0} we compare this short range order to the short range order  at $T=0$ in the phase where $\frac{J_2}{|J_1|}=0$.

We obtain the magnitudes of \J and \jj as follows. The experimental value of \tjt \@ is known, \tjt \@ $\simeq 750K$, and from Fig. \ref{Cv}, \tjt \@ $= 750 = \alpha J_2$ where $\alpha$ is a constant which is obtained from the slope, $J_2\simeq0.12 eV$. We find \J from the condition $0.05<\frac{J_2}{|J_1|}<0.25$, which gives $J_1$, $1.4eV>|J_1|>0.3eV$, which includes the value obtained by Millis {\it et al}.\cite{Millis96}, $0.6 eV$.

\begin{figure}[hi]
\includegraphics[scale=0.7]{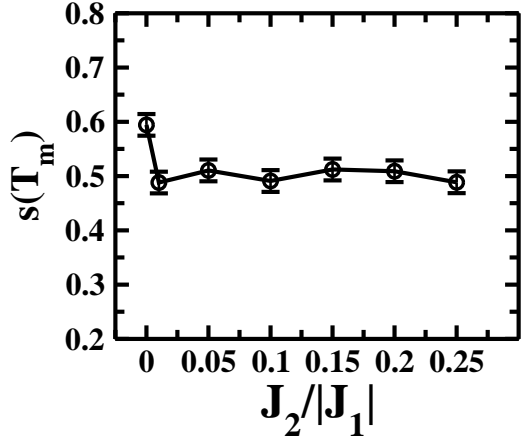}
\caption{\label{sph} The entropy for \lmo phase versus $J_2/|J_1|$ above \tjt \@ where \jj = 0.0, 0.01, 0.05 and 0.1 and \J$<0$.}
\end{figure}

The specific heat curves have a minimum at a temperature, $T_m$, between the two peaks. We evaluate the entropy at  $T_m$ by integrating the specific heat obtained from the simulation above $T_m$ and a temperature $T_H>>J_1/k_B$ where the entropy is given by $k_Blog_e3$.
\begin{equation}
k_Blog_e3 -s(T_m)=\int^{\infty}_{T_m}\frac{C_v(T)}{T}dT.
\end{equation}
The values of $s(T_m)$ obtained in this way are plotted in Fig. \ref{sph} as a function of \jj/$J_1$.

Figure \ref{sph} shows that $s(T_m)$ values are almost independent of $J_2/|J_1|$ along the two-peak region. The average value $s(T_m)=(0.50\pm0.02)k_B$ is in good agreement with the experimental value obtained by S\'anchez {\it et al.} who studied the local structure of \lmo across the JT transition at \tjt \@ $= 750K$ by means of X-ray absorption near edge structure and extended X-ray absorption fine structure at the Mn $K$-edge. They obtained from the heat capacity measurements of \lmo that the change in entropy between low $T$ and above \tjt, which is given as $s(T_m)$ in our theory, is given by $(0.515\pm0.02)k_B$. 

Since the relevant Potts model has $J_2/|J_1|$ small it is instructive to consider, in next section, the limiting case of \jj = 0.

\vspace{0.5cm}
\begin{figure}[hi]
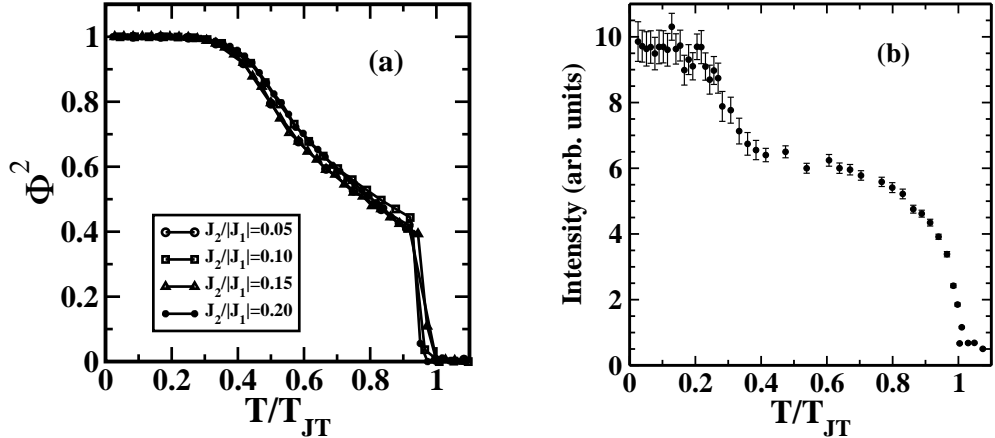

\includegraphics[scale=0.45]{OP_test.eps}\hspace{1.0cm} \includegraphics[scale=0.485]{murakami.eps} 
\caption{\label{OP}(a) The square of the order parameter, $\Phi^2$, of the orbital ordering in \lmo vs $T/T_{JT}$ at $J_2/|J_1|=0.05, 0.10,15$ and 0.20 with $L=12$, (b) The experimental results for the square of the order parameter of the orbital ordering as function of the reduced temperature, $T$/\tjt, (after Murakami {\it et al.}\cite{Murakami98}).}
\end{figure}
The square of the order parameter of the orbital ordering, $\Phi^2$, in \lmo is found as a function of the reduced temperature, $\frac{T}{T_{JT}}$, from the simulation as shown in Fig. \ref{OP}. We obtain the same value of $\Phi^2(T/T_{JT})$ from Monte Carlo simulations independent of the direction of temperature change taking account of the fact that there are three nontrivial inequivalent ordered phases, $|\Phi|^2$, corresponding to a choice of ${\bf Q}$. 

For ordering characterised by ${\bf Q}=(110)$ we define a site (n,m,l) as belonging to the odd (even) sublattice if $n+m$ is an odd (even) integer. We define the probability, $p^{(o)}_x$, by defining the total number of odd sites that are occupied by an $x$ orbital, $N_x^o$.
  \begin{equation}
  p^{(o)}_x =\frac{2N_x^o}{L^3}
  \end{equation}
This definition is extended to define $p^o_y$ and $p^o_z$ and similarly for the even sites $p^e_x$, $p^e_y$ and $p^e_z$. The order parameter of the orbital ordering for an $x$ or $y$ orbit in an odd site is, 
 \begin{equation}
\phi^{(o)}_x =p^{(o)}_x-\frac{1}{2}(p^{(o)}_y+p^{(o)}_z),
\end{equation}
and
\begin{equation}
\phi^{(o)}_y =p^{(o)}_y-\frac{1}{2}(p^{(o)}_z+p^{(o)}_x).
\end{equation}
Similarly, we are able to obtain the ordering on the even sites (when $m+n$ is even) e.g.,
\begin{equation}
\phi^{(e)}_x = p^{(e)}_x - \frac{1}{2} (p^{(e)}_y+p^{(e)}_z),
 \end{equation}
where $p^{(e)}_x, p^{(e)}_y$ and $p^{(e)}_z$ are the probability to find the even site occupied by $x, y$ or $z$ orbit respectively. The total order parameter of the orbital ordering in \lmo is,
\begin{equation}
\Phi_{(110)}(T/T_{JT}) =\frac{(\phi_x + \phi_y)}{2},
\end{equation}
where $\phi_x=\phi^{(o)}_x-\phi^{(e)}_x$ and $\phi_y=\phi^{(e)}_y-\phi^{(o)}_y$. We do not have a term for $\phi_z$ because it is not included in $\Phi_{(110)}(T/T_{JT})$ when ${\bf Q}=(110)$. When we start the simulation from high-$T$, we find that the system goes into one of the three equivalent ${\bf Q}$ ordered phases, (110), (101) or (011) at the same value of \tjt. 

Figure \ref{OP}(a) shows that the square of the order parameter of orbital ordering in this model phase has an unusual behaviour. The orbital ordering is essentially perfect for $T\le 0.4$\tjt. It decreases through the range $0.4<\frac{T}{T_{JT}}<1$ faster than  an Ising model. Finally, it goes rapidly to zero at \tjt \@ because there is phase transition at this temperature which appears to be first order (but this is not proved rigorously because of the finite size of the simulation). The order parameter is plotted as a function of the reduced temperature, $T/T_{JT}$, in Fig. \ref{OP}(a). It may be seen that the order parameter is remarkably insensitive to the value of \jj/\J provided that $J_2/J_1<<1$. Figure \ref{OP}(b) shows the experimental results from Murakami {\it et al.}\cite{Murakami98}. There is a qualitative agreement between the simple model proposed here and the experiments. The size of the first order jump at \tjt \@ is well reproduced by the Potts model. The experimental results have a strong concave region for $0.25\lesssim\frac {T} {T_{JT}} \lesssim0.9$ which is reproduced qualitatively by the theory. This is unusual behaviour of an order parameter as in most cases (for example Ising model) the value of $|\frac{dM}{dT}|$ increases monotonically with temperature.

\section{\label{J2=0}Phase for \jj = 0}
In this section we show how a study of the phase obtained for $J_2/|J_1| = 0$ gives further understanding of the orbital ordering in \lmo just above \tjt. When we set \J as AF and put \jj = 0 (at \jj = 0 there is no phase transition) a phase which has some physics of \lmo phase above \tjt \@ is obtained. Namely, there is no longer side-to side interaction but the head-to-head configuration is still disfavoured. 

The energy varies from the high-$T$ limit, $u(\infty)=J_1/3$ to zero at $T=0$ where there are no head-to-head configurations. In comparison of the new model with AF Potts model, because the anisotropic case has fewer antiferromagnetic interaction we expect higher value for the ground state entropy, $s(0)$, at $T=0$.

Therefore, the configuration of \lmo above \tjt \@ is affected by the short range order obtained by the head-to-head interaction. It is known that the entropy of the 3-state Potts model is $k_Blog3$ at high-$T$ and nearly $\frac{k_B}{2}log_e2$ at $T=0$.\cite{Banavar82} The anisotropic Potts model has an entropy at $T=0$ equal to $s(0)=(0.6\pm0.02)k_B$ which is dramatically higher than $s(T_m)$ that for \lmo phase above \tjt \@ through the range $0<\frac{J_2}{|J_1|}<0.25$.

\section{Conclusion}
A modified model called the anisotropic Potts model has been presented to study the orbital ordering and the local JT distortion above JT temperature, \tjt, in \lmo. The interactions between the three possible orbits depended both on the type of orbits and their relative direction on the lattice. Suitable values of the exchange interaction have been chosen to give the correct low temperature ground state for the orbital ordering phase in LaMnO$_3$. The phase obtained from these interaction values has a first order transition at the JT temperature, \tjt. 

Short range order was obtained above \tjt \@ for a wide ratio of the exchange interactions $0<J_2/|J_1|<0.25$. The short range order reduced the entropy from the value of three randomly occupied orbitals, $k_Blog_e3=1.098k_B$, to be $(0.5\pm0.02)k_B$ which is in fair agreement with the experimental value obtained by S\'anchez {\it et al}.\cite{Sanchez03}. When the side-to-side interaction was put to zero, \jj = 0, we obtained a phase that does not order down to the lowest temperature. This phase had an entropy at low temperature that was higher than that obtained for \lmo phase above \tjt. 

We obtained the order parameter of the orbital ordering phases as a function of temperature in good agreement with the published results. The square of the order parameter, $\Phi^2(T/T_{JT})$, of the orbital ordering decreases faster for $T>0.4/T_{JT}$ than an Ising model. At \tjt \@ it goes to zero discontinuously in a first order transition. 

The ratio of \jj/\J necessary to give the \lmo ground state are $0<J_2/|J_1|\le0.25$ and within this range we find that the results are almost independent of the exact value of \jj/$|J_1|$. In particular we find similar shapes of the curve $\Phi^2$ as a function of the reduced temperature, $T/$\tjt, as shown in Fig. \ref{OP} and also that entropy above \tjt \@ as shown in Fig. \ref{sph} is also approximately constant. Furthermore, the range of values of \jj/$|J_1|$ required to give the observed ordering temperature is consistent with those proposed from a microscopic model. Thus we have shown that the temperature dependence of the orbital ordering in \lmo is represented well by a phenomenological three state Potts model.

\begin{acknowledgements}
The authors acknowledge useful discussions with Prof. D. I. Khomskii  and are also indebted to Prof. Youichi Murakami for reading this manuscript and providing us with the experimental data for figure \ref{OP}b. This work is funded by the Egyptian High Education Ministry (EHEM).
\end{acknowledgements}

   \end{document}